\documentstyle[amssymb,prb,aps,twocolumn]{revtex}

\begin{document}
\date{\today}
\draft
\title{Spin-accumulation in small ferromagnetic double barrier junctions}
\author{Arne Brataas,$^{1,2}$ Yu. V. Nazarov,$^{1}$ J. Inoue$^{1,3}$ and Gerrit E.W.
Bauer$^{1}$}
\address{$^{1}$Laboratory of Applied Physics and \\
Delft Institute of Microelectronics and Submicrontechnology (DIMES), \\
Delft University of Technology, 2628 CJ Delft, The Netherlands}
\address{$^{2}$Philips Research Laboratories, Prof. Holstlaan 4, 5656 AA Eindhoven,\\
The Netherlands}
\address{$^{3}$Nagoya University, Department of Applied Physics, Nagoya, Aichi\\
464-01, Japan}
\maketitle

\begin{abstract}
The non-equilibrium spin accumulation in ferromagnetic double barrier
junctions is shown to govern the transport in small structures. Transport
properties of such systems are described by a generalization of the theory
of the Coulomb blockade. The spin accumulation enhances the
magnetoresistance. The transient non-linear transport properties are
predicted to provide a unique experimental evidence of the spin-accumulation
in the form of a reversed current on time scales of the order of the
spin-flip relaxation time.
\end{abstract}


In the seventies it was understood that electron transport in tunneling and
heterostructures involving metallic ferromagnets is associated with
non-equilibrium spins.\cite{Tedrow94:174,Johnson85:1790} Compared to other
time scales in electron transport the spin relaxation time is generally very
long at low temperatures, being limited only by scattering at paramagnetic
impurities and by spin-orbit scattering. The spin-relaxation time and the
spin-diffusion length which govern the spin accumulation has been measured
by Johnson in polycrystalline gold films.\cite{Johnson93:2142} The concept
of non-equilibrium spin accumulation plays an important role in the
Boltzmann theory of transport of the giant magnetoresistance in the current
perpendicular to the plane (CPP) configuration .\cite{Valet93,Gijs97}
However, the experimental evidence for the spin accumulation is indirect at
best. It can be shown that in the linear response regime the spin- and
charge- distribution functions can be completely integrated out of the
transport problem, which then depends exclusively on the scattering
probabilities and the applied bias.\cite{Gijs97} In this Rapid Communication
we show theoretically how unambiguous evidence for a non-equilibrium spin
accumulation can be obtained by the DC and AC response of ferromagnetic
double barrier junctions in the non-linear regime. These junctions have to
be small in order to observe big effects, which means that the complications
of the Coulomb blockade have to be taken into account (for a review see Ref.
[\onlinecite{Grabert92}]). To this end we have to extend very recent
theories of the Coulomb blockade in ferromagnetic double barrier junctions 
\cite{Barnas98:1058} to include time dependence and a non-zero spin
relaxation time.\cite{Brataas98} Ono et al. succeeded in fabricating a
ferromagnetic single electron transistor,\cite{Ono96:3449} which in
principle can be used to test our predictions. Coulomb charging effects have
also been seen in discontinuous multilayers\cite{Sankar97:5512} and in small
cobalt clusters.\cite{Schelp97:R5747}

We first show that the spin accumulation in a ferromagnetic double barrier
junctions becomes relevant when the number of electrons in the island
between the tunneling barriers is relatively small. In ferromagnetic
structures where the tunneling rates depend on the electron spin, a finite
current through the system is accompanied by a spin current out of or into
the island $(\partial s/\partial t)_{\text{tr}}$. This creates a
non-equilibrium excess spin $s$ on the island, which decays with the
spin-flip relaxation time $\tau _{\text{sf}}$ so that in steady state $%
(\partial s/\partial t)_{\text{tr}}=s/\tau _{\text{sf}}$. Energy relaxation
is much faster than spin relaxation, so that the occupation of the states
for each spin direction can be described by Fermi distributions.\cite
{Grabert92} The non-equilibrium spin accumulation on the island is
equivalent to a chemical potential difference $\Delta \mu $ between the spin
up and the spin down states. Since spin relaxation is slow and the
structures of interest are small, $\Delta \mu $ is uniform over a
sufficiently small island. In terms of the typical single-particle energy
spacing (or inverse energy density of states at the Fermi energy) $\delta $
we have $\Delta \mu =s\delta $. Spin accumulation may be expected to
interfere with the transport properties when $\Delta \mu $ is of the same
order as the applied voltage $V$. The spin current is of the same order as
the current, $e(\partial s/\partial t)_{\text{tr}}\thicksim I\thicksim V/R$,
where $R$ is the typical junction resistance. The non-equilibrium spin
accumulation is therefore important when the spin-relaxation time and/or the
single-particle energy spacing are sufficiently large: 
\begin{equation}
\tau _{\text{sf}}\delta /h>R/R_{K},  \label{condition}
\end{equation}
where the quantum resistance is $R_{K}=h/e^{2}$. The spin-flip relaxation
time in polycrystalline aluminium is $\tau _{\text{sf}}\sim 10^{-10}s$\cite
{Tedrow94:174} ($10^{-8}s$ in single-crystal aluminium at $T=4.3K$\cite
{Johnson85:1790} ) and $\tau _{\text{sf}}\thicksim 10^{-11}s$ for gold.\cite
{Johnson93:2142} The single-particle energy spacing on the island is roughly 
$\delta \thicksim E_{F}/N$, where $N$ is the number of atoms on the island
and $E_{F}\thicksim 10eV$ is the Fermi energy. In an Al island with less
than 10$^{6}$ atoms (10$^{8}$ atoms in single-crystals) the spin
accumulation may therefore be expected to play a significant role.
''Modern'' metals, like arm-chair nanotubes\cite{Tans98:49} or (magnetic)
semiconductor heterostructures\cite{Matsukara98:R2037} can also be
interesting as island materials. The first because of a possible huge
spin-flip relaxation time and the latter since islands containing a small
number of electrons can be created by depletion of the two-dimensional
electron gas.\cite{Grabert92}

In small systems where Eq. (\ref{condition}) is satisfied the spin-flip
relaxation time is longer than the charge relaxation time $RC$ ($C$ is the
capacitance of the island). This can be seen from Eq. (\ref{condition}), $%
\tau _{\text{sf}}>(2E_{c}/\delta )\cdot RC$, and noting that the charging
energy is larger than the single-particle energy spacing except in
few-electron systems, $E_{C}/\delta \sim \left( e^{2}/E_{F}a\right) N^{2/3}$
($e^{2}/E_{F}a\sim 1$). Hence the long-time response of the system is
dominated by the spin dynamics.

We consider a normal metal island attached to two ferromagnetic leads by two
tunnel junctions. We assume collinear magnetizations in the leads and
disregard size quantization. The tunnel junctions are characterized by a
capacitance $C_{i}$ and magnetic configuration-dependent conductances $%
G_{i\sigma },$ where $i=1,2$ denotes the first and the second junction and $%
\sigma $ denotes up $(+)$ or down $(-)$ spin electrons on the island. There
is a source-drain voltage $V$ between the right and the left reservoir and a
gate voltage source coupled capacitively to the island. Here we consider the
situation with a maximum Coulomb gap where the offset charge controlled by
the gate voltage is zero.\cite{Brataas98}

We proceed from the assumptions of the orthodox theory, i.e. $G_{i\sigma
}<G_{K}$ neglecting co-tunneling,\cite{Takahashi98:1758} with the difference
that the transition rates becomes spin-dependent. The transition rate from
the left reservoir to the island is 
\begin{equation}
\overrightarrow{\Gamma ^{1\sigma }}_{n+1,n}=\frac{1}{e^{2}}G_{1\sigma
}F(E_{1}(V,q)-\sigma \Delta \mu /2),  \label{rate}
\end{equation}
where the energy difference associated with the tunneling of one electron
into the island through junction $i$ is\cite{Grabert92} $E_{i}(V,q)=\kappa
_{i}eV+e(q-e/2)/(C_{1}+C_{2})$, the charge on the island is $q=-ne$, the
total capacitance is $1/C=1/C_{1}+1/C_{2}$, $\kappa _{i}=C/C_{i}$, $%
F(E)=E/[1-\exp (-E/k_{B}T)]$ and $k_{B}T$ is the thermal energy. The spin
balance is 
\begin{equation}
\frac{ds}{dt}=\left( \frac{ds}{dt}\right) _{\text{tr}}+\left( \frac{ds}{dt}%
\right) _{\text{ rel}},  \label{spineq}
\end{equation}
where the spin-relaxation rate is $\left( ds/dt\right) _{\text{rel}}=-s/\tau
_{\text{sf}}=-\Delta \mu /\delta \tau _{\text{sf}}$, $\tau _{\text{sf}}$ is
the spin-flip relaxation time and $\delta ^{-1}$ is the density of states at
the Fermi level in the island. The spin balance (\ref{spineq}) can be
written in the stationary case as $I_{s}=e(ds/dt)_{\text{tr}}=G_{s}2\Delta
\mu /e$, where the ''spin relaxation conductance'' is introduced as $%
G_{s}\equiv e^{2}/2\delta \tau _{\text{sf}}$. The master equation\cite
{Grabert92} determines the probability $p_{n}$ to have $n$ excess electrons
on the island. The current through the first junction is $%
I_{1}=(I_{1}^{\uparrow }+I_{1}^{\downarrow })$, where the current of
electrons with spin $\sigma $ is $I_{1}^{\sigma }=e\sum_{n}p_{n}(%
\overrightarrow{\Gamma ^{1\sigma }}_{n+1,n}-\overleftarrow{\Gamma ^{1\sigma }%
}_{n-1,n})$ and there is a similar expression for the current through the
second junction $I_{2}=(I_{2}^{\uparrow }+I_{2}^{\downarrow })$. The spin
current is 
\begin{equation}
\left( \frac{ds}{dt}\right) _{\text{tr}}=(I_{1}^{\uparrow
}-I_{1}^{\downarrow }-I_{2}^{\uparrow }+I_{2}^{\downarrow })/e.
\label{spintr}
\end{equation}
In the Coulomb blockade regime the current is zero, $I=0$, and it can be
shown that $\Delta \mu $ vanishes, as expected.\cite{Brataas98} The Coulomb
gap in the low-temperature current-voltage characteristics is thus not
modified by the non-equilibrium spin accumulation. We also want to point out
that for symmetric tunneling junctions $G_{1\uparrow }/G_{1\downarrow
}=G_{2\uparrow }/G_{2\downarrow }$ the non-equilibrium spin accumulation
vanishes and our theory reduces to those in Refs. [\onlinecite{Barnas98:1058}%
].

In this orthodox model the problem can be mapped on the equivalent circuit
in Fig.\ \ref{f:cir} by introducing the ''spin capacitance'' $C_{s}\equiv
e^{2}/2\delta $, so that 
\[
(es)/2=C_{s}(\Delta \mu /e)\text{, }\Delta \mu /s=e^{2}/(2C_{s})=\delta . 
\]
This ''charging energy'' of the spin capacitance is thus simply the
single-particle energy cost of a spin-flip, $\delta $, or more generally,
the inverse of the magnetic susceptibility $\mu _{B}^{2}/\chi _{s}$.

We solve the general problem for the steady state as well as for the
time-dependent properties by numerically integrating the master equation and
the spin balance, Eq.\ (\ref{spineq}). We choose symmetric capacitances $%
C_{1}=C_{2}=C$ in our calculations. Thus the important energy scale is the
Coulomb energy $E_{c}=e^{2}/2C$ and the other relevant energies are
renormalized by $E_{c}$. The thermal energy is $k_{B}T=0.05E_{c}$. The
spin-dependent junction conductances are described in units of the average
junction conductance $G$ and the currents are normalized by $Ge/2C$. In the
parallel configuration, the conductances are $G_{1\sigma }^{\text{P}%
}=G_{1}(1+\sigma P)/2$ and $G_{2\sigma }^{\text{P}}=G_{2}(1+\sigma P)/2$,
where $P$ is the polarization of the ferromagnets. In the antiparallel
configuration $G_{1\sigma }^{\text{AP}}=G_{1}(1+\sigma P)/2$ and $G_{2\sigma
}^{\text{AP}}=G_{2}(1-\sigma P)/2$.

We consider first the steady state transport properties where the spin
capacitance $C_{s}$ does not contribute. The junction magnetoresistance is
the relative difference in the resistance when switching from the
antiparallel to the parallel configuration. In the absence of the
non-equilibrium spin accumulation, the junction magnetoresistance vanishes
for the F/N/F junction. The spin accumulation causes a non-zero
magnetoresistance. We show in Fig. \ref{f:tmr} the calculated junction
magnetoresistance for $G_{1}/G=1$, $G_{2}/G=2$ and a polarization $P=0.4$ in
the limit of slow spin-relaxation $G_{s}/G=0$ (upper curve) and fast
spin-relaxation $G_{s}/G=5$ (lower curve). We see the magnetoresistance
oscillations as a function of the source-drain voltage.\cite{Barnas98:1058}
The amplitude of the oscillations decreases with increasing source-drain
voltage, where the Coulomb charging is less important.\cite{Barnas98:1058}
The period of the oscillations is close to $2E_{c}$ for our system. There is
only a small distortion of the shape of the magnetoresistance oscillations
with increasing spin-relaxation rate in the island. The magnetoresistance
and its oscillations are noticeable even when the spin-relaxation
conductance is of the same order as the tunnel conductances, in agreement
with Eq. (\ref{condition}). In the absence of the Coulomb charging energy,
the tunnel magnetoresistance is 
\begin{equation}
\text{TMR}=P^{2}\frac{1-\gamma ^{2}}{1-P^{2}\gamma ^{2}+\alpha ^{2}},
\label{JMRFNF}
\end{equation}
where $\gamma =(G_{1}-G_{2})/(G_{1}+G_{2})$ is a measure of the asymmetry of
the junction conductances and $\alpha ^{2}=4G_{s}/(G_{1}+G_{2})$ determines
the reduction of the magnetoresistance due to the spin-relaxation. For a
high source-drain bias when the Coulomb charging effects are negligible, the
numerical results agree well with Eq. (\ref{JMRFNF}), TMR$=11\%$ for $%
G_{s}/G=0$ and TMR$=2\%$ for $G_{s}/G=5$.

For the transient response in the antiparallel configuration we use $P=0.5$, 
$G_{1}/G=1.3$, $G_{2}/G=2.6$ and $G_{s}/G=0.3$. Let us consider first a
fixed source-drain voltage at a high bias until the system is stationary and
then lower the source-drain voltage. We have used $\tau _{\text{sf }}=10RC$
(e.g. $E_{c}=0.2$ meV and $R/R_{K}=10$ gives $RC=2\cdot 10^{-11}s$). The
initial high bias is $V_{i}=10E_{c}$ which gives a stationary current of $%
I_{i}=6.2Ge/2C$ and we investigate the behavior of the transient current
when the final source-drain bias is below, $V_{f}=0$ ($I_{f}=0$), and above
the Coulomb charging energy, $V_{f}=4E_{c}$ ($I_{f}=2.1Ge/2C$). We show in
the upper panel in Fig.\ \ref{f:cur} the current through the first and the
second junction for $V_{f}=4E_{c}$ (upper curves) and $V_{f}=0$ (lower
curves) after the source-drain voltage is changed at $t=0$. It is clearly
seen that the relaxation of the current is slow on the time scale $RC$. For
time scales less than $RC$, we see that the current through the first and
the second junction are not the same due to the charge depopulation in the
island. The average of the upper curves ($V_{f}=4E_{c}$) where the final
source-drain voltage is well above the Coulomb blockade energy, follows to
within 10-20\% the description given by the equivalent circuit neglecting
the Coulomb charging effects described below (\ref{imp}), (\ref{chem}) and (%
\ref{stime}) according to which the spin accumulation time is $\tau _{\text{%
spin}}=2.4RC$. When the source-drain voltage is switched off ($V_{f}=0$), we
see that the transient current is negative. However, the spin accumulation
time is much longer in this case, $\tau _{\text{spin}}\simeq \tau _{\text{sf}%
}=10RC$. This discrepancy becomes more evident when we consider the relative
change of $\Delta \mu $ or $s$: 
\[
D(t)\equiv \left| \frac{s(t=\infty )-s(t)}{s(t=\infty )-s(t=0)}\right| . 
\]
In the lower panel of Fig.\ \ref{f:cur} we show the calculated
time-dependent relative change $D(t)$ in the situations $V_{f}=4E_{c}$
(upper solid curve) and $V_{f}=0$ (lower solid curve), which are found to be
remarkably different.

In order to understand the dynamics it is useful to inspect the device
without the Coulomb charging effects, i.e. the capacitances $C_{1}$ and $%
C_{2}$ in the equivalent electric circuit in Fig.\ \ref{f:cir}. We set the
voltage on the left lead to zero and apply a time dependent potential $V(t)$
to the right lead. The complex impedance $Z_{\text{spin}}(\omega )=V(\omega
)/I(\omega )$ is 
\begin{equation}
\frac{1}{Z_{\text{spin}}(\omega )}=\frac{G_{1}G_{2}}{G_{1}+G_{2}}-\frac{%
G_{1\uparrow }G_{2\downarrow }-G_{1\downarrow }G_{2\uparrow }}{(G_{1}+G_{2})}%
\frac{\Delta \mu (\omega )}{eV(\omega )}  \label{imp}
\end{equation}
where 
\begin{equation}
\frac{\Delta \mu (\omega )}{eV(\omega )}=\frac{1}{1+i\omega \tau _{\text{spin%
}}}\frac{G_{1\uparrow }G_{2\downarrow }-G_{1\downarrow }G_{2\uparrow }}{%
(G_{s}+G^{\prime })(G_{1}+G_{2})}.  \label{chem}
\end{equation}
Here the spin accumulation time is 
\begin{equation}
\tau _{\text{spin}}=\frac{C_{s}}{G_{s}+G^{\prime }},  \label{stime}
\end{equation}
where $1/G^{\prime }=1/(G_{1\uparrow }+G_{2\uparrow })+1/(G_{1\downarrow
}+G_{2\downarrow })$. From the relations (\ref{imp}) and (\ref{chem}) we see
why switching-off the source-drain voltage ($V_{f}=0$) reverses the
transient current as found in the upper panel in Fig.\ref{f:cur}. Without
the Coulomb blockade this transient decays on the time scale $\tau _{\text{%
spin}}$. In the limit that the junction conductances are much smaller than
the spin conductance, the spin accumulation time (\ref{stime}) reduces to
the spin-flip relaxation time, $\tau _{\text{spin}}\approx \tau _{\text{sf}}$
. In the opposite limit where the junction conductances are much larger than
the spin conductance, the spin accumulation time is $\tau _{\text{spin}}\sim
C_{s}R$. The spin-capacitance is much larger than the charge-capacitance $C$
in the regime where the orthodox theory is valid ($\delta \ll E_{C}$) and
thus the spin accumulation time is much larger than the charge-relaxation
time.

The dashed lines in the lower panel in Fig. \ref{f:cur} correspond to the
spin accumulation time in the absence of charging, $\tau _{\text{spin}%
}=2.4RC $ as well as to the spin-flip relaxation time $\tau _{\text{sf}%
}=10RC $. We see that the calculated spin accumulation time agrees well with
the equivalent circuit described above (\ref{stime}) for $V_{f}=4E_{c}$, but
disagrees with this expression for $V_{f}=0$ where the spin accumulation
time is close to $\tau _{\text{sf}}$. The latter is a result of the Coulomb
charging which is seen to affect the spin accumulation time. In this case
the non-equilibrium spin accumulation decays slower since the spins must
relax through the spin-conductance $G_{s}$ on the island and the transport
through the junctions is suppressed. In this situation, the relaxation time
of the non-equilibrium spins and the current for long times is equal to the
spin-flip relaxation time $\tau _{\text{sf}}$, as observed in Fig.\ \ref
{f:cur}.

It should be noted that the magnon assisted inelastic tunneling, which
reduces the TMR, gives negligibly small contribution in our case because of
a magnon excitation gap, presumably due to magnetic anisotropy and/or size
effects.\cite{Zhang97:3744} This magnon gap is larger than the bias voltage
applied in our study. For very small islands like the metallic cluster
studied in Ref. [\onlinecite{Schelp97:R5747}] when the Coulomb charging
energy is larger than the magnon gap, magnon inelastic tunneling can
interfere with the Coulomb charging effects.

In conclusion, \label{s:con}we have investigated the influence of a
non-equilibrium spin accumulation on the transport properties of a
ferromagnetic single-electron transistor. For a F/N/F junction we find a
finite magnetoresistance due to the non-equilibrium spin accumulation. The
spin accumulation can have a drastic effect on the transient transport
properties. A transient response can be found on time scales much larger
than the charge relaxation time $RC$. The same slow response is also
expected if other external parameters such as the gate voltage, or the
magnetization are changed.

This work is part of the research program for the ``Stichting voor
Fundamenteel Onderzoek der Materie'' (FOM), which is financially supported
by the ''Nederlandse Organisatie voor Wetenschappelijk Onderzoek'' (NWO). We
acknowledge benefits from the TMR Research Network on ``Interface
Magnetism'' under contract No. FMRX-CT96-0089 (DG12-MIHT) and the ``Monbusho
International Scientific Research Program on Transport and Magnetism of
Microfabricated Magnets''. G. E. W. B. would like to thank Seigo Tarucha and
his group members for their hospitality at the NTT Basic Research
Laboratories and Keiji Ono for a discussion.

\begin{figure}[tbp]
\caption{The equivalent circuit for the current-voltage response of the
system.}
\label{f:cir}
\end{figure}

\begin{figure}[tbp]
\caption{The junction magnetoresistance in the limit of no spin-relaxation
in the island ($G_S/G=0$) and fast spin relaxation ($G_S/G=5$).}
\label{f:tmr}
\end{figure}

\begin{figure}[tbp]
\caption{The current as a function of time (upper panel). The relative
change of the non-equilibrium spin as a function of time (lower panel). The
source-drain voltage is switched from $V_i=10E_c$ to $V_f=4E_c$ or $V_f=0$
at $t=0$.}
\label{f:cur}
\end{figure}

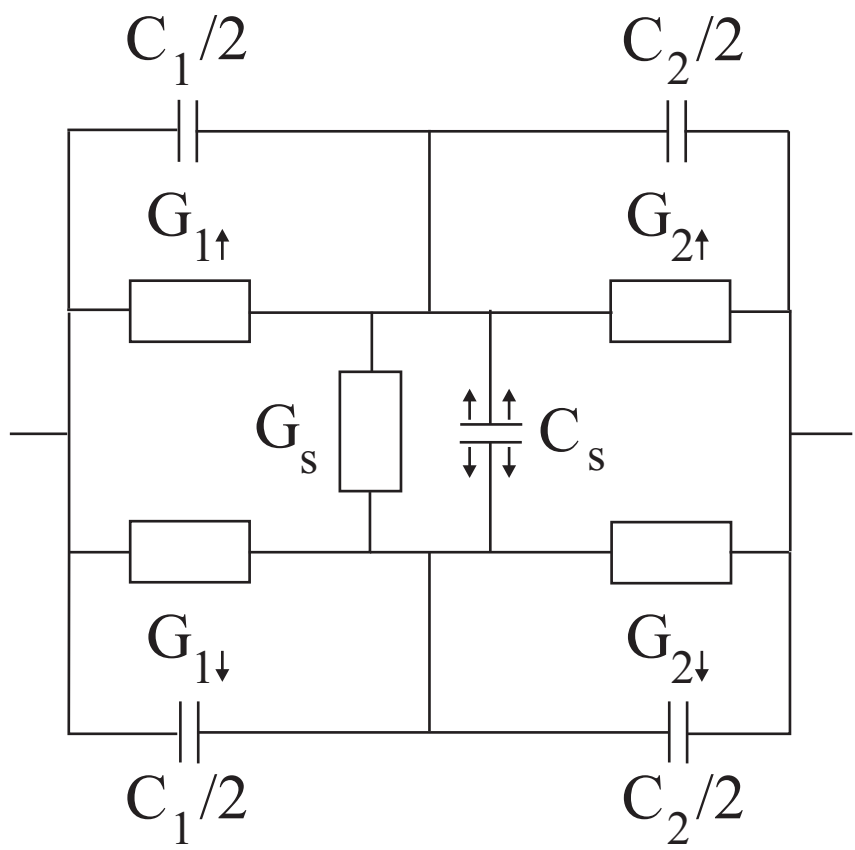
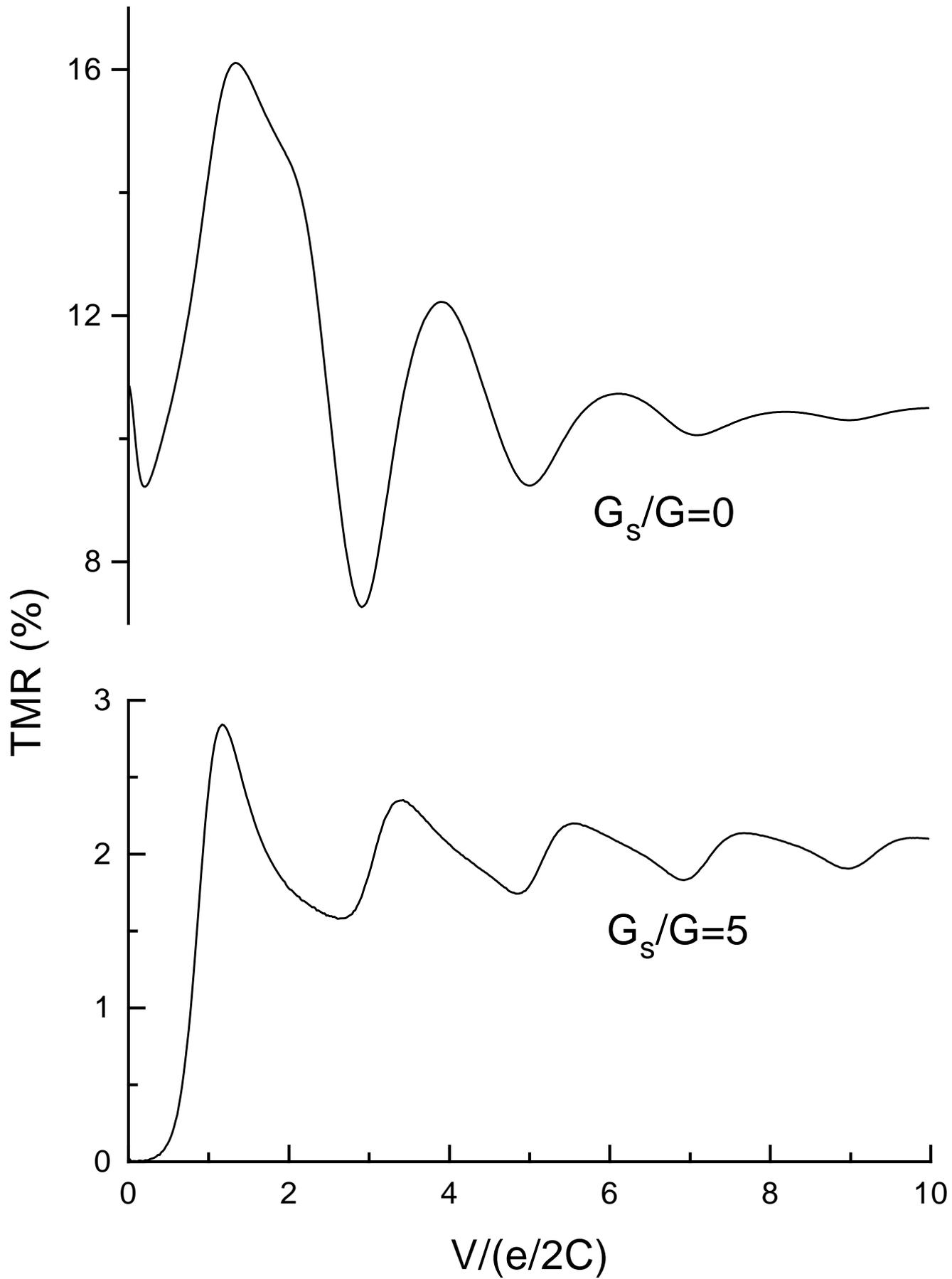
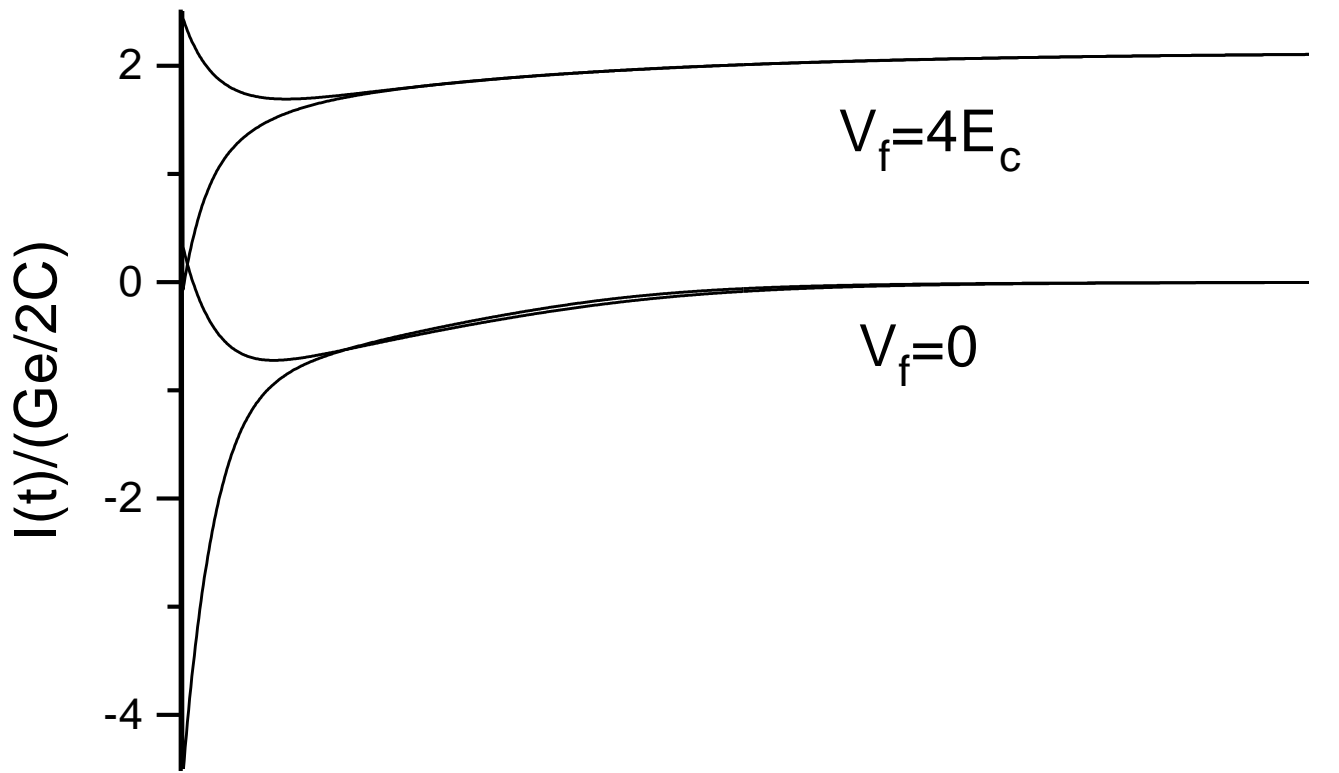
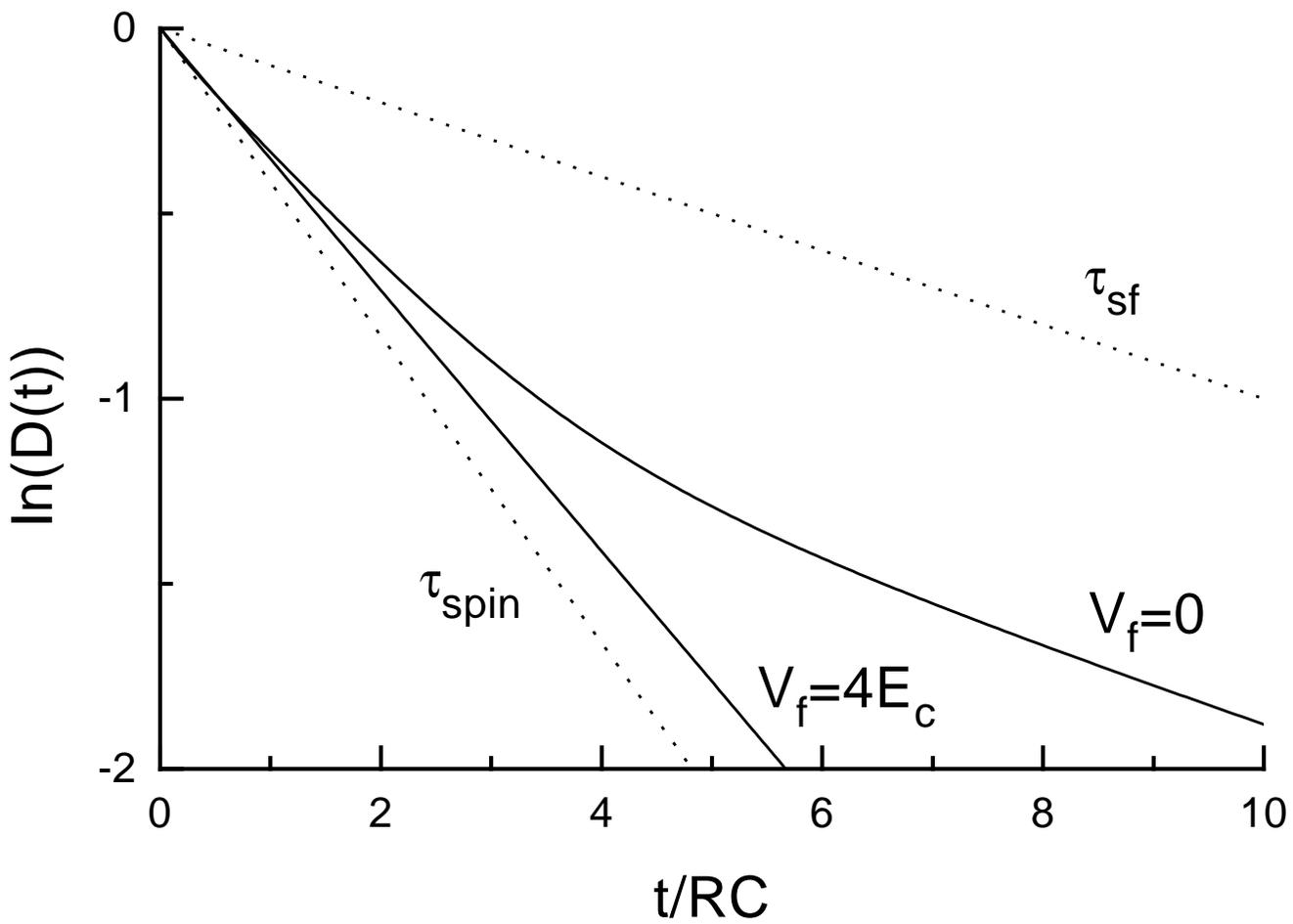
\end{document}